# Freeway Lane Management Approach in Mixed Traffic Environment with Connected Autonomous Vehicles

Omar Hussain, Amir Ghiasi, Xiaopeng Li

Department of Civil and Environmental Engineering, University of South Florida, Tampa, FL 33620, USA

**Abstract**
Connected autonomous vehicles (CAV) technologies are about to be in the market in the near future. This requires transportation facilities ready to operate in a mixed traffic environment where a portion of vehicles are CAVs and the remaining are manual vehicles. Since CAVs are able to run with less spacing and headway compared with manual vehicles or mixed traffic, allocating a number of freeway lanes exclusive to CAVs may improve the overall performance of freeways. In this paper, we propose an analytical managed lane model to evaluate the freeway flow in mixed traffic and to determine the optimal number of lanes to be allocated to CAVs. The proposed model is investigated in two different operation environments: single-lane and managed lane environments. We further define three different CAV technology scenarios: neutral, conservative, and aggressive. In the single-lane problem, the influence of CAV penetration rates on mixed traffic capacity is examined in each scenario. In the managed lanes problem, we propose a method to determine the optimal number of dedicated lanes for CAVs under different settings. A number of numerical examples with different geometries and demand levels are investigated for all three scenarios. A sensitivity analysis on the penetration rates is conducted. The results show that more aggressive CAV technologies need less specific allocated lanes because they can follow the vehicles with less time and space headways.

*Keywords*: Connected automated vehicles; managed lane; mixed traffic; headway; optimization

## Introduction
Operations of connected automated vehicles (CAVs) have recently drawn intensive attention from researchers in transportation engineering, particularly on investigating the potential benefits CAVs will bring in terms of mobility, safety, and environmental impacts. These benefits are actually largely determined by the operational characteristics of future CAV technologies, that yet have quite some uncertainties. For example, autonomous vehicle (AV) technologies (e.g., adaptive cruise control) allow may not largely impact vehicle headway. However, if AVs are connected and form CAVs, their headway can be largely reduced since a following vehicle knows its preceding vehicle's intended movements in real time with connected vehicle communications *(1)*. An example is Cooperative Adaptive Cruise Control (CACC), which can shorten the following vehicles' reaction times in response to the leading vehicle's actions and consequentially improving traffic throughput and decreasing fuel consumption and emissions *(2)*. The headway characteristics of mixed traffic including both CAVs and manual vehicles are even more complicated. Manual vehicle's tends to keep a roughly 2 second's headway when following another manual vehicle *(3)*, which may be further subject to various uncertainties *(4)*. While following an autonomous vehicle, human being's behavior may be different from following a manual vehicle, since the driver may have different expectations of vehicle dynamics between a manual vehicle and an autonomous vehicle. Vice versa, when CAV follows a manual vehicle, due to the lack of communication, its response might be quite distinguished from following another CAV. Such different characteristics of headways in mixed traffic may largely affect traffic dynamic and system capacity.

Managed lane (ML) strategies are being applied to freeways to improve the freeway performance, including mobility, travel time, travel speed, traffic flow, fuel consumption, safety, and congestion reduction. Managed lane is defined by the Federal Highway Administration (FHWA) as "freeway-within-a-freeway where a set of lanes within the freeway cross section is separated from the general-purpose lanes". From operational perspective, it is also defined by FHWA as "highway facilities or a set of lanes where operational strategies are proactively implemented and managed in response to changing conditions". Operational strategies could be categorized by the objectives to be achieved. The ML strategies' objectives implemented in today's highways include pricing, access control, vehicle eligibility, or their combinations *(5)*. In this study, we consider the eligibility strategy as our main objective. The most commonly deployed eligibility strategy in freeways is High-Occupancy Vehicles (HOV) lanes. In such strategy, one or more lanes are dedicated to HOVs in order to improve the person throughput, safety, travel time reliability, congestion reduction in the freeways *(6)*.

We propose to transfer the concept of ML into the CAV context. This idea extends the advantage of traditional ML to separation of CAVs from mixed traffic. Since the headway between two CAVs may be much shorter than that other types of headway, such lane separation, if properly designed, shall improve the overall traffic throughput and reduce bottleneck congestion. Although a number of recent studies mentioned the importance of integrating CAVs into ML strategies (e.g. *7–11*) , there is a lack of modeling techniques on how to optimally design the number of exclusive CAV lanes based on mixed traffic headway characteristics.

This study aims to bridge this modeling gap. The main contribution of this study is to develop an analytical model that determines the optimal number of lanes that should be allocated to CAVs in a freeway such that the freeway overall throughput is maximized. In this study, we consider the operation of CAVs and human-driven vehicles in mixed and separated traffic and evaluate the facility performance in both conditions. The process includes separating all or a portion of CAVs from human-driven vehicles by allocating a certain number of exclusive CAV lanes at which the freeway total throughput is maximized. A dynamic implementation of such strategy would allow for a better lane allocation in accordance with CAVs.

**Literature review**
In recent literature, increasing attention is given for modeling automated vehicles operation in mixed or separated traffic to improve traffic performance. Ma et al. *(12)* proposed a trajectory control algorithm, called a parsimonious Shooting Heuristic (SH) algorithm that controls trajectories of automated vehicles in an advanced highway section. Trajectories are constructed based on "the boundary condition for vehicle arrivals, vehicle mechanical limits, traffic lights and vehicle following safety". They concluded that the proposed trajectory control has a positive impact on improving overall traffic performance. In other studies, a couple of variable speed limit (VSL) control models and algorithms are proposed for automated vehicles on a busy connected isolated environment. The basic concept of these models is controlling automated vehicles speeds in a way that improves facility operation. Benefits achieved from the previous VSL models are bottleneck discharge rate improvement and system delay reduction *(13)*, as well as total travel time and fuel consumption reductions and safety improvements *(14)*. Similarly, Wang et al. *(15)* presented a VSL model in a mixed environment with an additional vehicle acceleration control. The system output benefits are traffic efficiency and sustainability improvements (e.g., stop-and-go waves solutions). In terms of advanced models, Gu, et al. *(16)* developed an adaptive control model for the leading vehicle of a Cooperative Adaptive Cruise Control platoon in mixed traffic flows. The model controls the leading vehicle actions by a minimization optimization problem for reaction acceleration of the following vehicles in that platoon. The developed model was able to improve traffic flow stability through CACC platoons. In the same field, Roncoli et al. *(17)* developed a predictive control framework that controls actions of automated vehicles ACC. The framework includes a quadratic optimization problem for ramp metering, vehicle speed, and lane changing controls. As for managing freeway merge conflicts, Park et al. *(18)* presented an advanced freeway Merging Control Algorithm that controls both gaps of ramp vehicles and "lead & lag" gaps of mainline vehicles so that vehicles in the ramp experience smoother merge. Simulation results of the control model show increases in vehicles miles traveled and average speeds, and reduction in vehicles total travel times, vehicles total delay times, and fuel consumption. In the same field of ramp metering control, Weg et al. *(19)* proposed advanced cooperative systems that eliminate stop-and-go behaviors near on-ramps by controlling vehicles' speeds in the mainline during and after the moving jam. The most important finding emphasized in this study is the ability of the proposed systems to improve a freeway throughput.

Furthermore, a number of recent studies has been dedicated to ML operations and lane allocation to automated vehicles. Fakharian Qom et al. *(20)* estimates the impacts of introducing and prioritizing automated vehicles on ML by setting ML a pricing schedule and access restrictions. VISSIM software package was used to estimate the performance of the managed and general-purpose lanes through considering a macroscopic traffic model based on Static Traffic Assignment (STA) and mesoscopic simulations based on Dynamic Traffic Assignment (DTA). The model results show that the increase in penetration of automated vehicles using ML contributes to ML pricing schedule and reduction in the GP lanes congestion. STA and DTA output trends are consistent; however, DTA shows more increase in automated vehicles penetration on ML due to its better modeling of congestion. Su et al. *(21)* has also considered automated vehicles operation on ML. However, this study proposes an improved Lane-Changing assistant system that allows the automated vehicles to adjust their speeds in a ML as to provide a sufficient gap for a vehicle in GP lanes that requested a lane-change. The system was developed to always maintain the speed differential between Managed and GP lanes. An experiment was tested on a two-way freeway segment with one lane for HOV and the other for GP lane. where the penetration of vehicles switching from GP lanes to ML was predetermined. As for lanes dedication, Hearne and Siddiqui *(22)* has presented a discussion study in which most of the operational and safety aspects of lane allocation for automated vehicles were covered, including definition of dedicated lanes, throughput on dedicated lanes, impact of dedicated lanes on GP lanes, and look-ahead in the right-of-way and interchange issues. However, few studies modeled how to optimally determine the number of CAV lanes that will optimize the performance of mixed traffic. This study aims to address this issue.

**Research goal and objectives**
The main objective of this research is to develop a model to determine the optimal number of lanes to be allocated to CAVs. Basically, if the freeway flow in mixed traffic is at or exceeds capacity, the separation decision may allow the freeway to discharge at higher throughput. Thus, we determine the number of CAV lanes that maximizes throughput under congested states. To achieve this goal, we plan to
- Formulate the average headway for traffic with mixed CAV and manual vehicles at different CAV penetration rates.

- Develop an analytical model to evaluate the freeway overall throughput under different lane management decisions and to determine the optimal CAV lane number.
- Conduct a sensitivity analysis in ML operations to observe the effects of CAVs penetration rates on the optimal number of CAV lanes.
- Evaluate the developed model in different scenarios by varying the CAVs penetration rates, total number of the freeway lanes, and total demand values on the freeway overall throughput.

**Problem statement**
*Single Lane Problem*
This section describes a single-lane problem. We consider a one-lane freeway without any inflow or outflow ramps as illustrated in Figure 1. In this problem, we consider a mixed traffic environment that CAVs and manual vehicles are randomly distributed. We assume that all CAVs are identical and all manual vehicles have statistically identical behavior (e.g., identical average headways). This leads to generation of four types of headways as illustrated in Figure 1. Each class of vehicles comprises two average headways. For a CAV, it follows either another CAV or a manual vehicle, which we denote as $\bar{h}_{AA}$ and $\bar{h}_{AM}$, respectively. Similarly, for a manual vehicle, we denote $\bar{h}_{MM}$ and $\bar{h}_{MA}$ respectively as the average headway for a manual vehicle following another manual vehicle, and the average headway for a manual vehicle following CAV.
The average headway of vehicles in mixed traffic can be calculated as:

$$\bar{h}_{mix} = \frac{N_{AA}\bar{h}_{AA} + N_{AM}\bar{h}_{AM} + N_{MM}\bar{h}_{MM} + N_{MA}\bar{h}_{MA}}{N_{AA} + N_{AM} + N_{MM} + N_{MA}} \quad (1)$$

Where,
$N_{AA}$: Number of the CAVs following other CAVs.
$N_{AM}$: Number of the CAVs following manual vehicles.
$N_{MA}$: Number of the manual vehicles following CAVs.
$N_{MM}$: Number of the manual vehicles following other manual vehicles.

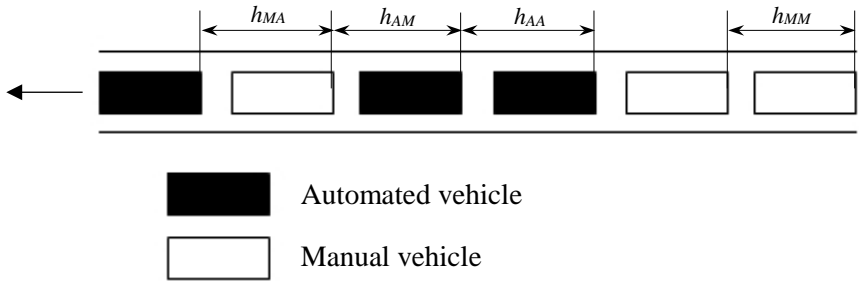

**FIGURE 1 Types of Headways in a Single-Lane Mixed Traffic**

Note that the single lane traffic over a long period of time can be regarded as a ring road. Then the total number of CAVs and manual vehicles are $N_A = N_{AA} + N_{AM}$ and $N_M = N_{MM} + N_{MA}$, respectively. As we consider manual vehicles and CAVs are randomly distributed, and since $N_A = q_{mix}P_A$ and $N_M = q_{mix}(1 - P_A)$, the average headway of vehicles in mixed traffic can be expressed as a function of penetrations of automated vehicles as follows:

$$\bar{h}_{mix}(P_A) = P_A^2 \bar{h}_{AA} + P_A(1-P_A)(\bar{h}_{AM} + \bar{h}_{MA}) + (1-P_A)^2 \bar{h}_{MM} \qquad (2)$$

Where, $P_A$ is the total penetration rate of CAVs in mixed traffic. Thus, the mixed traffic capacity for a one-lane freeway can be calculated as:

$$c_{mix} = \frac{1}{\bar{h}_{mix}} \qquad (3)$$

*Managed Lane Problem*
In this section we intend to determine the optimal number of CAV lanes, denoted by $l_A$, such that the freeway throughput is maximized. With this, $L - l_A$ lanes of the freeway will be allocated to mixed traffic, where $L$ is the total number of freeway lanes. In ML problem we assume that $L \geq 2$. The ML objective function is to maximize overall throughput of the freeway that is calculated as:

$$Q = (L - l_A) q_{mix} + l_A q_A \qquad (4)$$

Where,
$Q$: The total traffic throughput (veh/hr),
$q_{mix}$: The mixed traffic throughput in any of the $L - l_A$ lanes (veh/hr/lane),
$q_A$: The purely automated traffic throughput in any of the $l_A$ lanes (veh/hr/lane),
$C_{mix}$: The capacity at any of the unallocated lanes (veh/hr/lane),
$C_A$: The capacity at any of the allocated lanes (veh/hr/lane).

Note that since on allocated lanes, there is not any manual vehicle, and CAVs can maintain the minimum headway at any traffic status, we assume that $C_A$ is a fixed value. $q_A$ is in fact a function of $P_A$, $D$ and $l_A$ that is calculated as

$$q_A(P_A, D, l_A) = \frac{\min(P_A L D, l_A C_A)}{\max(1, l_A)} \qquad (5)$$

Where $D$ is the total per lane demand.

$C_{mix}$ mainly relies on the penetration rate of CAVs left in unallocated lanes (i.e., $1, \ldots, L - l_A$), which we denote as $p'_A$. When $q_A > C_A$, a portion of total CAVs will be directed to the $L - l_A$ unallocated lanes to generate mixed traffic with manual vehicles. Otherwise, $p'_A = 0$. $p'_A$ can be determined as

$$p'_A = \frac{\max(0, (P_A L D - l_A C_A))}{\max(1, L D - l_A q_A)} \qquad (6)$$

Thus, $C_{mix}$ can be defined as a function of $p'_A$ as follows:

$$C_{mix}(p'_A) = \frac{1}{\bar{h}(p'_A)} = \frac{1}{{p'_A}^2 \bar{h}_{AA} + p'_A(1-p'_A)(\bar{h}_{AM} + \bar{h}_{MA}) + (1-p'_A)^2 \bar{h}_{MM}} \qquad (7)$$

We calculate $q_{mix}$ in the unallocated lanes a function of $P_A, D$ and $l_A$ as

$$q_{mix}(P_A, D, l_A) = \min\left(\frac{LD - l_A q_A}{\max(1, L - l_A)}, C_{mix}(p'_A)\right) \qquad (8)$$

As mentioned before, the ML problem aims to find the optimal number for $l_A$ such that $Q$ is maximized. Note that in equation (4), $Q$ is defined as a function of $l_A$, $q_{mix}$ and $q_A$, and since $l_A$ is an integer value, the proposed optimization problem is an integer programming (IP) that is expressed as:

$$\underset{l_A}{\text{Maximize}} \quad Q = (L - l_A)\, q_{mix} + l_A\, q_A \qquad (9)$$
$$\text{Subject to} \quad l_A \in [0, 1, \cdots, L],$$

Since the feasibility region of the proposed optimization problem is a set with $L + 1$ elements, it efficiently can be solved by the exhaustive enumeration method. In other words, we evaluate all possible values of $l_A$ (i.e., $\forall l_A = 0, \ldots, L$) and select the one that maximizes the objective.

**Numerical analysis**
In this section, we perform a number of numerical experiments to assess the performance of the proposed optimization problem. In these numerical examples, without the loss of generality, we assume that $h_{AM} = h_{MA}$. We define three scenarios for CAV technologies by varying $h_{AA}, h_{AM}$ (or $h_{MA}$) and $h_{MM}$. These three scenarios are based on three different CAV technologies: conservative, neutral and aggressive. The three scenarios considered are as follows:
1. Neutral ($h_{AA} < h_{AM} < h_{MM}$), where $h_{AA} = 0.45\ s$, $h_{AM} = 1.2\ s$, and $h_{MM} = 1.8\ s$
2. Conservative ($h_{AA} < h_{AM} = h_{MM}$), where $h_{AA} = 0.45\ s$, $h_{AM} = 1.8\ s$, and $h_{MM} = 1.8\ s$
3. Aggressive ($h_{AA} <<< h_{AM} < h_{MM}$), where $h_{AA} = 0.3\ s$, $h_{AM} = 1.2\ s$, and $h_{MM} = 1.8\ s$

The first scenario is the base condition in which we define three normal headways as mentioned above. In the second scenario, we assume that CAVs are a bit more conservative when they follow a human-driven vehicle, so in this scenario we increase $h_{AM}$ and set an equal number with $h_{MM}$ as mentioned above. In the third scenario, we assume that the CAVs maintain lower headways when they follow each other. In this case, we decrease $h_{AA}$ to a lower value than the two other scenarios.

In this section, we first consider a single-lane problem to examine the influence of $P_A$ changes on the mixed traffic capacity (i.e., $c_{mix}$). Further, we provide a number of examples with different $L$ and $D$ values for the ML problem that are followed by some sensitivity analyses on $P_A$.

*Single-Lane Problem*
In this section, we perform a sensitivity analysis on the influence of $P_A$ changes on $c_{mix}$ which is shown in Figure 2. The results show that as $P_A$ increases in the aggressive scenario, $c_{mix}$ can be increased up to 12000 vehicles per hour per lane (i.e., for $P_A = 1$); however, $c_{mix}$ in the neutral and the conservative scenarios can reach the value of 8000 vehicles per hour per lane (i.e., for $P_A = 1$). In addition, the influence of $P_A$ increase on $c_{mix}$ is higher in the aggressive scenario than in the neutral scenario, especially when the penetration rate exceeds 40%. In the conservative scenario, however, $P_A$ increase has less influence on $c_{mix}$, even at the lower penetrations rates.

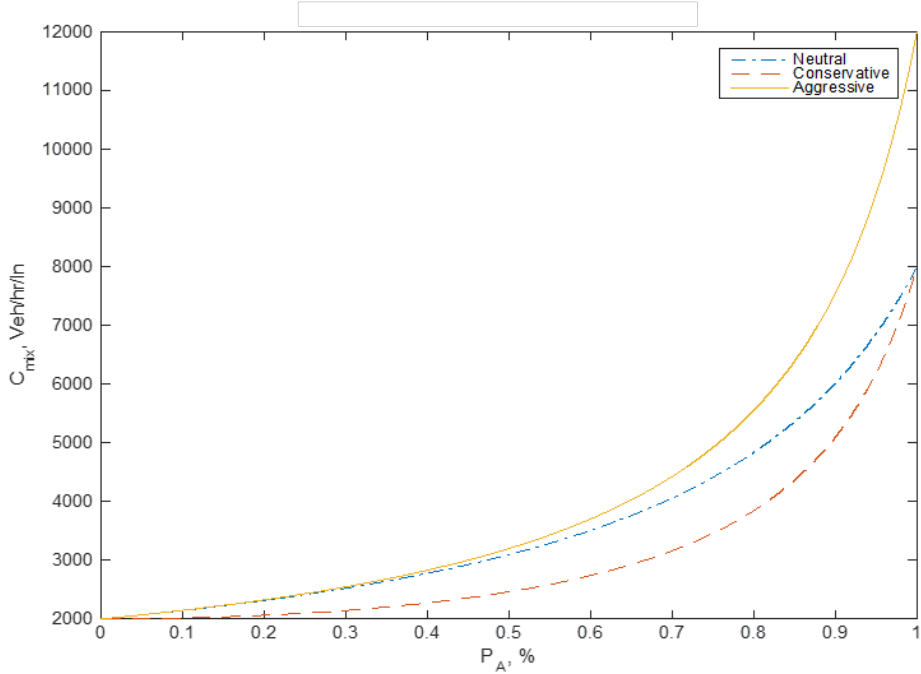

**FIGURE 2 Impact of $P_A$ on $C_{mix}$ (Single-Lane and $L$ = 2, 4, and 6, Capacity ≥ Demand)**

*Managed Lane Problem*
In this section, we analyze the effects of $P_A$ changes on $l_A$. We investigate the three aforementioned scenarios for different total number of lanes, i.e., $L$ = 2, 4 and 6. We also consider two cases of demand for each scenario. In the first case, $D$ is less than the total capacity, and for the second case, we assume that $D$ is greater than the total capacity. In what follows, we perform a sensitivity analysis of $P_A$ for the three CAV scenarios and for each combination of cases.
    $L$ = 2, 4, and 6, Capacity ≥ Demand: In this case, since D is less than the capacity, the traffic status is at free-flow conditions, and allocating any lane to CAVs will not change the total traffic throughput. Thus, there is no need to allocate any lane to CAVs. In such situations, the effect of $P_A$ changes on the capacity in all scenarios is the same as that of the single-lane problem that is previously discussed (Figure 2). Note that this fact applies to all L values.
    $L$ = 2, Capacity < Demand: When the demand exceeds the capacity, the freeway operate under congestion. In this case, allocating a number of lanes to CAVs may improve the total throughput. We perform a sensitivity analysis on $P_A$ to investigate the influence of its changes on the optimal number of $l_A$. In this case (i.e., $L$=2), we found that one lane will be allocated to CAVs at certain ranges of $P_A$ values, depending on the scenario. Figure 3 shows the result of this analysis. As shown in this figure, in the conservative scenario, one lane is allocated to CAVs at $0.21 \leq P_A \leq 0.90$. However, this range is $0.30 \leq P_A \leq 0.81$ and $0.31 \leq P_A \leq 0.72$ for the neutral and the aggressive scenarios, respectively. The differences are because of the fact that for a specific value of $P_A$, $C_{mix}$ is higher in the aggressive scenario. This implies that the need to allocate one lane for CAVs in the aggressive scenario emerges at higher values of $P_A$ than the other two scenarios. Moreover, as $P_A$ increases, $C_{mix}$ reaches higher values in the aggressive scenario, and the need to allocated one lane for CAVs eliminates at lower values of $P_A$. The same procedure

occurs for the neutral scenario compared to the conservative one. This idea can also be seen in Figure 4. As $P_A$ increases, $C_{mix}$ reaches higher values in the aggressive scenario except for the aforementioned $P_A$ ranges that correspond to a lane allocation for CAVs. In $P_A$ ranges that CAVs are in the allocated lane, $C_{mix}$ drops significantly to 2000 veh/hour/lane because there are no CAVs left in the unallocated lane, and this lane operates with only human-driven vehicles that causes capacity drop. However, this capacity drop is compensated by the throughput increase in the allocated lane (i.e., $q_A$) that is resulted in higher total throughput (i.e, $Q$).

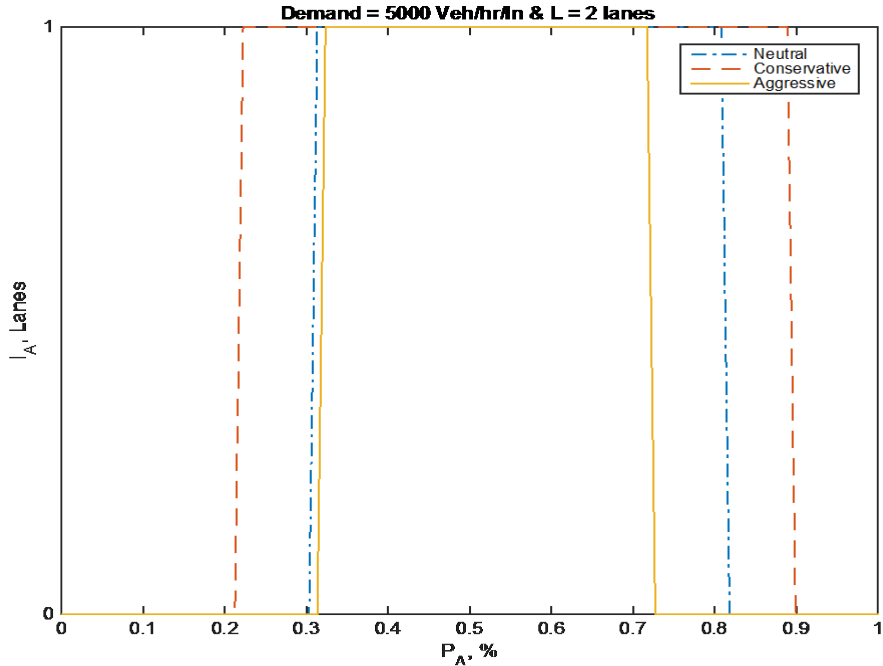

**FIGURE 3 Impact of $P_A$ Changes on $l_A$ ($L = 2$, Capacity < Demand)**

*L = 4, Capacity < Demand:* We perform similar analysis to assess the effect of $P_A$ changes on $l_A$ when $L = 4$ and the demand exceeds the capacity. The results are shown in Figure 5. It is found that two lanes are allocated to CAVs when $P_A$ is in the range of 57%-83% and 51%-89% for the neutral and the conservative scenarios, respectively. However, one allocated lane is sufficient when $P_A$ in the range of 14%-56% and 12%-50%, respectively. In the aggressive scenario, when $P_A$ is in the range of 14%-75%, only one lane is sufficient for CAVs. This is again because of the fact that CAVs can maintain lower headways in the aggressive scenario that results in higher $C_{mix}$. Thus, in this scenario, $l_A$ is less than that of the other two scenarios.

*L = 6, Capacity < Demand:* We repeat the same analysis for $L = 6$, and the results are shown in Figure 6. The results show that in the aggressive scenario, $l_A$ is sometimes one or even two less than that of the two other scenarios. Both the neutral and conservative scenarios allow CAVs to have three allocated lanes when $P_A$ is in the range of 65%-82% and 57%-89%, respectively. In these scenarios, two lanes are allocated to CAVs when $P_A$ is in the range of 38%-78% and 34%-59%, respectively. In the aggressive scenario, for $P_A$ range of 9% to 53%, only one lane need to be allocated to CAVs, and after that until 76%, there should be one more allocated lane to discharge at the maximum throughput.

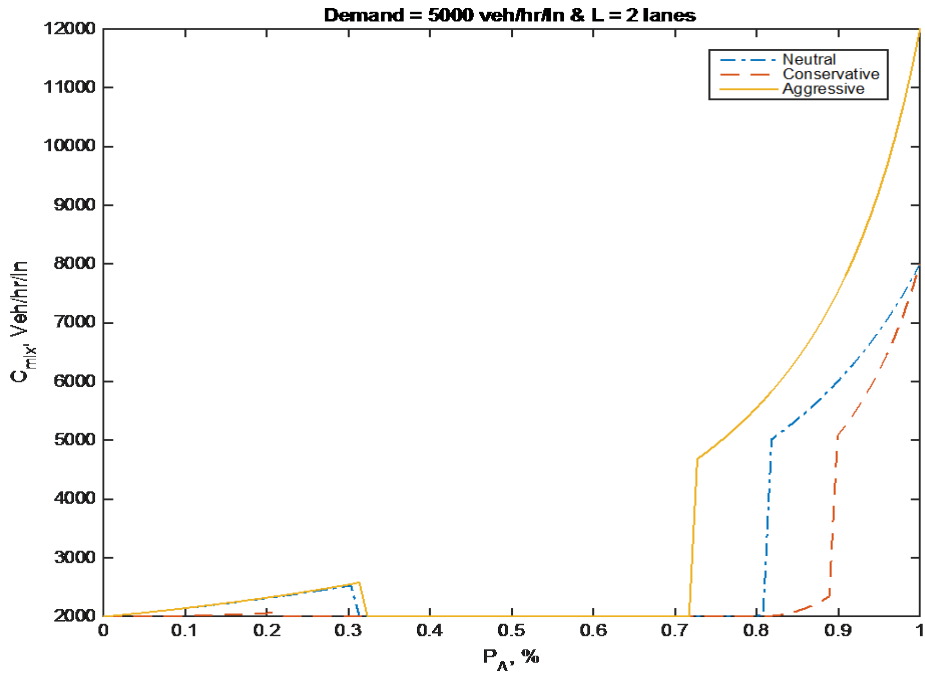

**FIGURE 4 Impact of $P_A$ Changes on $C_{mix}$ ($L = 2$, Capacity < Demand)**

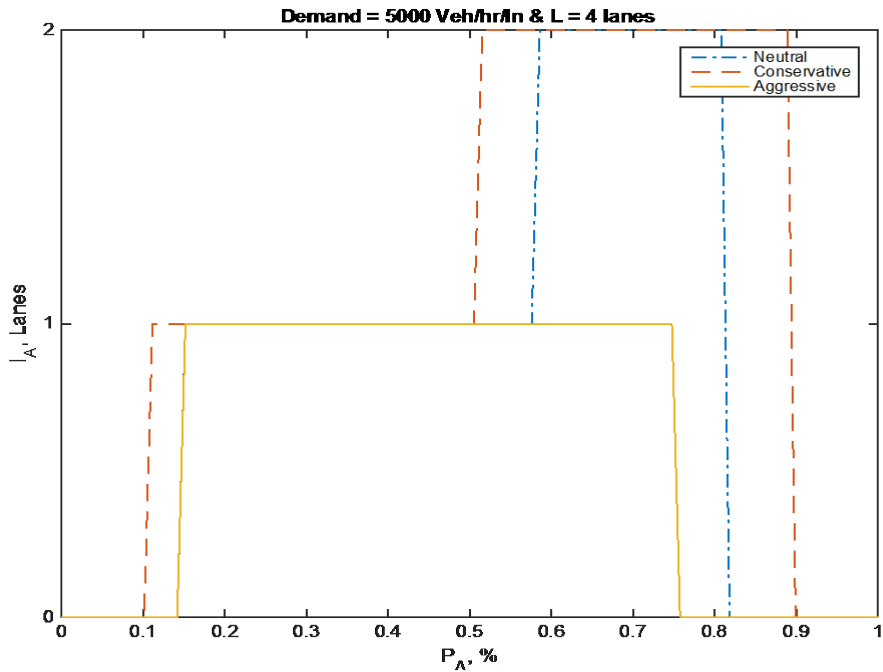

**FIGURE 5 Impact of $P_A$ Changes on $l_A$ ($L = 4$, Capacity < Demand)**

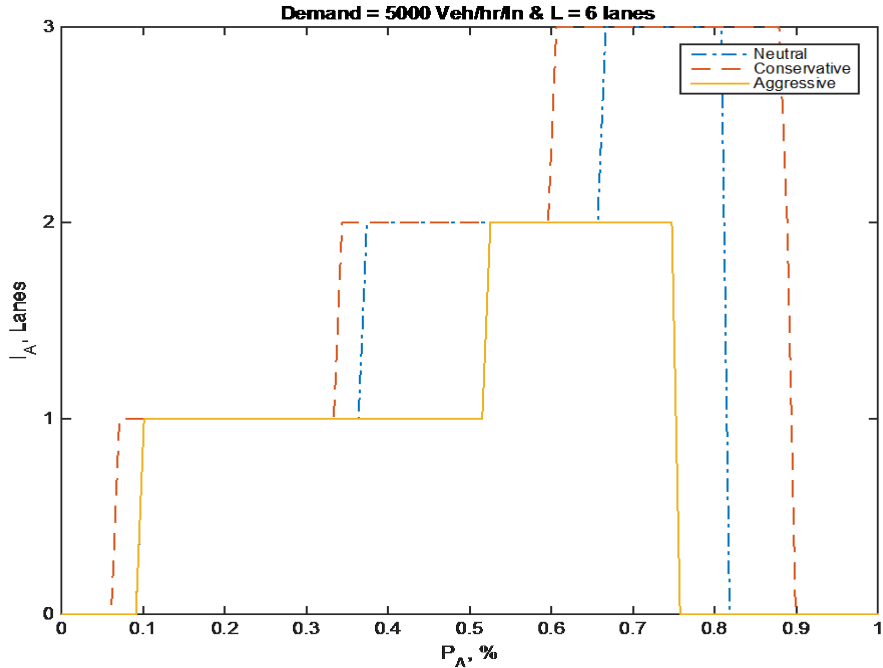

**FIGURE 6 Impact of $P_A$ Changes on $l_A$ ($L = 6$, Capacity < Demand)**

**Conclusions**
In this study, we propose an analytical managed lane model to determine the optimal number of lanes to be allocated to CAVs at which the freeway total throughput is maximized. To construct the optimization objective function, we first assume that the CAVs and human-driven vehicles are randomly distributed along a freeway. With this, we defined four types of headways that result in the mixed traffic capacity as a function of the total CAV penetration rate. Finally, the freeway total throughput is derived as a function of total CAVs penetration rate. The proposed optimization problem can be easily and efficiently solved by the exhaustive enumeration method.

A number of numerical examples are performed for the single-lane and ML problems. We defined three scenarios to differentiate three different CAVs technologies. These scenarios are called neutral, conservative and aggressive. For single-lane problem, numerical analysis show that as the CAV penetration rate increases, the freeway capacity increases as well. However, the increase achieved in the conservative scenario is not as high as those achieved in the neutral and aggressive scenarios. For ML problem, it is found that the decision on the lane allocation might take place only if the demand is greater than the freeway capacity. We performed some numerical analyses for different total number of lanes in the freeway, and the results show that in some CAV penetration rate ranges in the aggressive scenario, less lanes are allocated to CAVs than in the neutral or conservative scenarios. This findings show that more aggressive CAV technologies need less CAV allocated lanes because they can follow the vehicles with less time and space headways.

In this study, we considered the expected values of headways that are obtained from reviewing the relevant studies. However, headways in empirical observations shall follow statistical distributions. Considering statistical distributions for the headways will allow us to develop a stochastic ML problem that is considered for our future reaserch.

**Acknowledgements**

This research is partially supported by the Federal Highway Administration and Leidos, Inc. through Grant # DTFH61-12-00020 and the U.S. National Science Foundation through Grant CMMI#1453949.